\documentclass{iau}

\usepackage{amsmath}
\usepackage{graphicx}
\usepackage{multirow}

\newcommand{\fg}{\rm{1P}\ }

\newcommand{\sg}{\rm{2P}\ }

\begin{document}

\lefttitle{Mirek Giersz et al.}
\righttitle{Proceedings of the International Astronomical Union: No.398 and MODEST-25}

\jnlPage{1}{7}
\volno{398/MODEST-25}
\jnlDoiYr{2025}
\doival{10.1017/xxxxx}

\aopheadtitle{Proceedings IAU Symposium}
\editors{eds. Hyung Mok Lee, Rainer Spurzem and Jongsuk Hong}

\title{Simulations of Globular Cluster Evolution
with Multiple Stellar Populations}

\author{Mirek Giersz$^1$ \\
Abbas Askar$^1$, Arkadiusz Hypki$^{1,2}$, Jongsuk Hong$^3$, Grzegorz Wiktorowicz$^1$ and Lucas Hellström$^1$}
\affiliation{$^1$Nicolaus Copernicus Astronomical Centre, Polish Academy of Sciences, ul. Bartycka 18, 00-716 Warsaw, Poland, \email{email: mig@camk.edu.pl}}
\affiliation{$^2$Faculty of Mathematics and Computer Science, A. Mickiewicz University, Uniwersytetu Poznańskiego 4, 61-614 Poznań, Poland}
\affiliation{$^3$Korea Astronomy and Space Science Institute, Daejeon 34055, Republic of Korea}

\begin{abstract}
The formation of stars with light-element abundance variations in globular clusters and the subsequent dynamical evolution of these multiple populations remains an open question. One of the most widely discussed is the AGB scenario, in which chemically processed material from the envelopes of AGB stars mixes with re-accreted primordial gas flowing into the center of the cluster. Based on this scenario, more than two hundred MOCCA simulations of cluster evolution have been carried out, incorporating additional physical processes related to the external environment of globular clusters and the initial properties of multiple stellar populations. Analysis of the simulations shows that most observed properties of multiple stellar populations and the global parameters of Milky Way clusters are well reproduced, with the exception of the correlation between cluster mass and the fraction of second-population stars. We present a speculative scenario of globular cluster evolution that may account for the observed properties of Milky Way clusters, including the correlation between cluster mass and the fraction of enriched stars. The scenario further predicts that, under certain conditions, the pristine first population can be more centrally concentrated than the enriched second population, as observed in some clusters. 
\end{abstract}


\maketitle

\section{Introduction}\label{s:intro}
%\cite{Giersz2025a}
Since their discovery, globular clusters (GCs) have been regarded as the simplest dynamic systems with a spherically symmetric structure and uniform chemical composition. The age of globular clusters is comparable to that of the Universe, and their formation history is inextricably linked to the formation of galaxies. Over the past decades, photometric and spectroscopic studies have revealed the presence of multiple stellar populations (MSP), primarily characterized by internal variations in light-element abundances 
\cite[e.g.,][and references therein]{Bastian2018, Gratton2019}. 
Despite the great efforts of many theoretical and observational groups, the formation of MSPs in globular clusters is still a mystery. Observations, both spectroscopic and photometric, provide the following information:
\begin{itemize}
\item	In massive clusters, some stars have different abundances of light elements: C, O, sometimes Mg are depleted, and Y, N, Na, sometimes Al are enhanced. Abundances are correlated or anti-correlated. This indicates a high-temperature CNO cycle;
\item There is a very small Fe spread allowed (~0.1 dex);
\item There is practically no time spread between populations. At most about a hundred Myr;
\item Enriched population (2P) is centrally concentrated, but there are some exceptions;
\item MSPs are clearly detected in massive, larger than a few $10^4 M_{\odot}$, and old (older than about 2 Gyr) clusters, and are observed not only in the Milky Way (MW);
\item  Fraction of \sg increases with cluster mass, may increase with cluster age.
\end{itemize}
From a theoretical point of view, the above observational facts indicate that:
\begin{itemize}
\item The light-element abundance variations in MSPs most likely originate from hot hydrogen burning ($\approx 70$ MK), but not from helium burning;
\item Formation of MSP is connected with dense cluster environment – abundance trends essentially are not visible in the field population;
\item Gas pollution occurs practically at the time of cluster formation – very small observed age spread;
\item It seems that there are no special conditions needed in clusters, except cluster mass, to form MSP.
\end{itemize}
Several scenarios have been proposed to explain the emergence and continued evolution of MSPs. These models can be divided into two groups. The first group, in which the formation of an \sg is shifted in time relative to the pristine population (1P) and requires the re-accretion of the pristine gas by the GC and its mixing with matter ejected by Asymptotic Giant Branch (AGB) stars, belongs to the AGB scenario. The second group belongs to models in which \sg is formed at virtually the same time as \fg and does not require the re-accretion of gas by the GC. Matter ejected from massive stars and binary systems mixes with residual primordial gas. These scenarios include: interacting massive binaries, fast-rotating massive stars, early disk accretion, very massive stars, nucleosynthesis in accretion disks around stellar-mass black holes, and the recently renewed single-binary composite scenario. As pointed out in the review paper by \cite{Bastian2018}, no scenario can explain a significant number of observational facts. In this work, we focus on the AGB scenario, as it is the framework for the simulations performed with the MOCCA code. 

In the AGB scenario, once the residual gas from the formation of the \fg is expelled, the cluster remains near the original gas cloud. After some time, this gas is re-accreted onto the cluster, where it accumulates in the cluster center and mixes with material ejected from AGB star envelopes. This mixture is crucial for producing the observed light-element correlations and anti-correlations. Following a time delay, the \sg forms from the polluted gas accumulated in the cluster center.

In the next section, we present results from MOCCA simulations designed to test how environmental effects, gas re-accretion onto GCs, and the time delay of \sg formation influence the evolution of stellar populations. Detailed descriptions of the code and initial conditions are provided in \cite{Giersz2025a}. A total of about 200 simulations were performed, with most models starting as tidally filling (TF).

\section{Results}\label{s:result}

In this section, we analyze the impact of gas re-accretion, cluster migration, and some of the individual MSP and global cluster parameters on the evolution of the cluster mass, half-mass radius ($\mathrm{R_h}$), and ratio between the number of \sg stars to the total number of stars ($\mathrm{N_2/N_{tot}}$). 

\subsection{Gas re-accretion}\label{s:gas}
\begin{figure}[h!]
\includegraphics[width=1.0\textwidth, height=0.23\textheight]{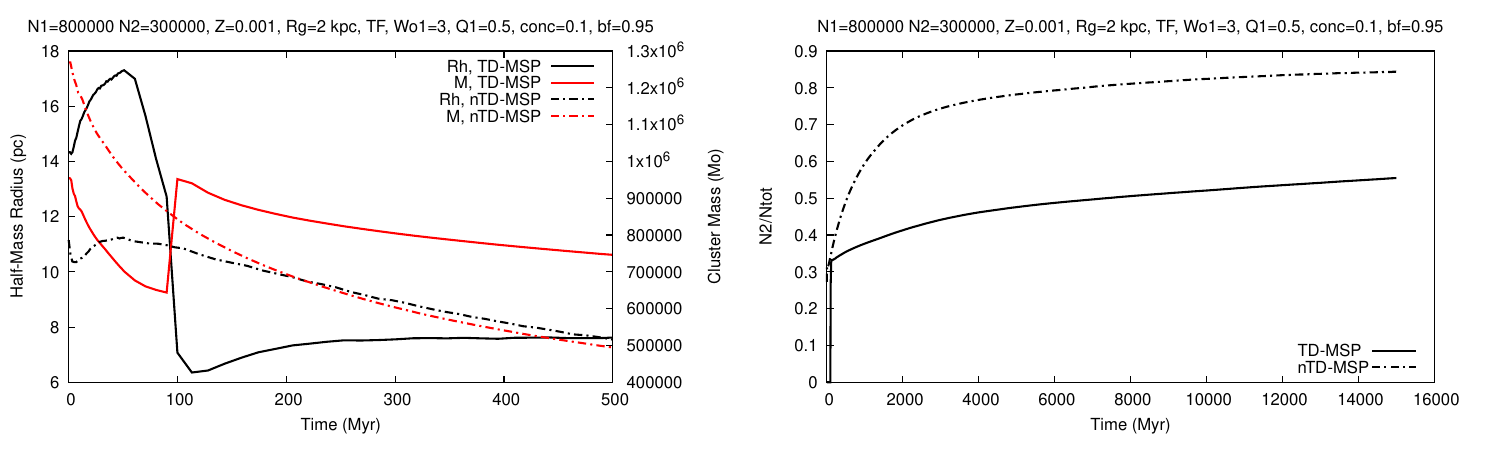} 
\caption{Left panel: Evolution of $\mathrm{R_h}$ for nTD-MSP model, and for TD-MSP model, and evolution of the total cluster mass for nTD-MSP model, and TD-MSP model. Right panel: Evolution of the ratio $\mathrm{N_2/N_{tot}}$ for the nTD-MSP model, and TD-MSP model. The global cluster parameters are listed at the top of each panel: $\mathrm{N_1}$ - number of \fg objects, $\mathrm{N_2}$ - number of \sg objects, $\mathrm{Z}$ - metallicity, $\mathrm{R_g}$ - galactocentric distance (size of the circular orbit), TF - tidally filling \fg, $\mathrm{W_{o1}}$ – King parameter for \fg, $\mathrm{conc}$ - concentration parameter ($\mathrm{conc_{pop}=R_{h2}/R_{h1}}$), $\mathrm{bf}$ – binary fraction, $\mathrm{Q_1}$ - virial ratio for \fg.}
\label{f:Fig1} 
\end{figure}

In the AGB scenario, re-accretion of gas on the cluster is required. We know that the ejection of gas from the cluster leads to a decrease in the cluster mass, cluster expansion, and half-mass radius growth. Thus, mass accretion should lead to an increase in the cluster's mass, its central concentration, and a decrease in the $\mathrm{R_h}$. The left panel of Figure~ \ref{f:Fig1} shows this behavior for the model with time delay formation of \sg stars (TD-MSP). In contrast, in the model where \sg stars form simultaneously with \fg (nTD-MSP), the evolution is driven only by stellar-evolution mass loss and star escapes caused by the tidal field. After gas re-accretion, the cluster collapses, transitioning from TF to under-filling (nTF). This transition leads to a significant slowdown in the growth of the $\mathrm{N_2/N_{tot}}$ ratio, as shown in the right panel of Figure~ \ref{f:Fig1}.

\subsection{Cluster migration}\label{s:migration}
\begin{figure}[h!]
\includegraphics[width=1.0\textwidth, height=0.23\textheight]{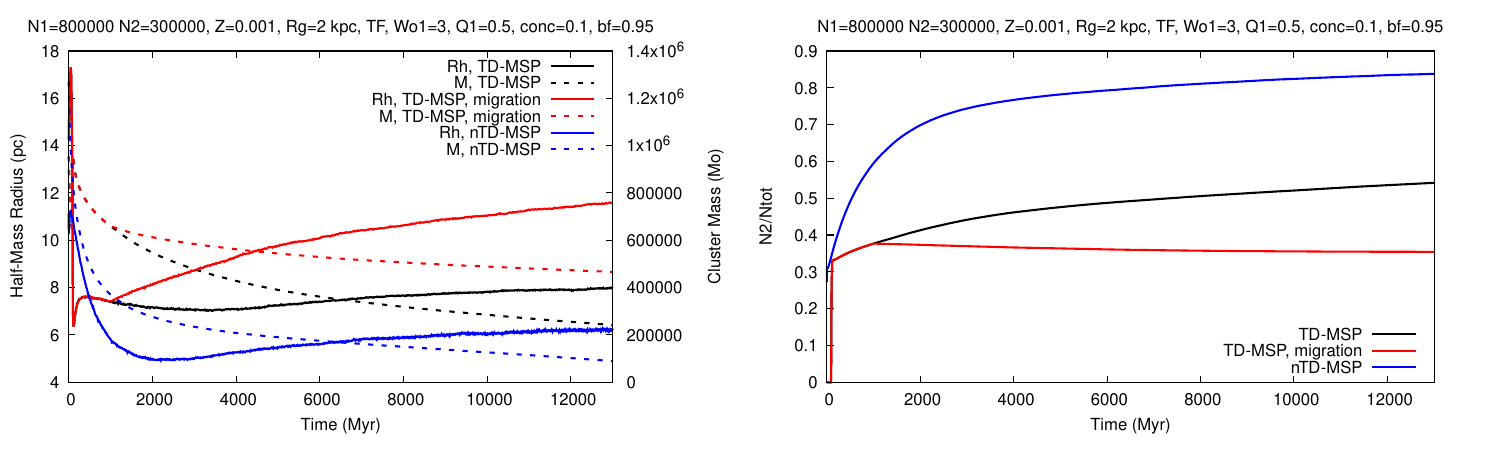} 
\caption{Left panel: Evolution of $\mathrm{R_h}$ for nTD-MSP model, TD-MSP model, and TD-MSP model with migration, and evolution of the total cluster mass for nTD-MSP model, TD-MSP model, and TD-MSP model with migration. Right panel: Evolution of the ratio $\mathrm{N_2/N_{tot}}$ for nTD-MSP model, TD-MSP model, and TD-MSP with migration model. The global cluster parameters are listed at the top of each panel and are the same as for Figure~\ref{f:Fig1}.}
\label{f:Fig2} 
\end{figure}

It is known from many papers, e.g. \cite{Meng2022} 
that as a result of strong changes in the tidal field of the forming galaxy, GCs can migrate further away from the center. Migration to greater distances means that clusters go from TF to nTF very quickly. This leads to a slower loss of cluster mass, an increase in half-mass radius, and virtually a halt in $\mathrm{N_2/N_{tot}}$ growth. Indeed, these predictions are confirmed in Figure~ \ref{f:Fig2}. The left panel shows a significant increase in $\mathrm{R_h}$ and a slower decrease in cluster mass with increasing $\mathrm{R_g}$. In the right panel, the ratio $\mathrm{N_2/N_{tot}}$ is practically constant after increasing the galactocentric distance. The migration of GCs introduces an additional parameter that influences cluster mass, $\mathrm{R_h}$, and the $\mathrm{N_2/N_{tot}}$ ratio.

\subsection{\fg virial ratio}\label{s:virial}
\begin{figure}[h!]
\centering
\includegraphics[width=1.0\textwidth, height=0.23\textheight]{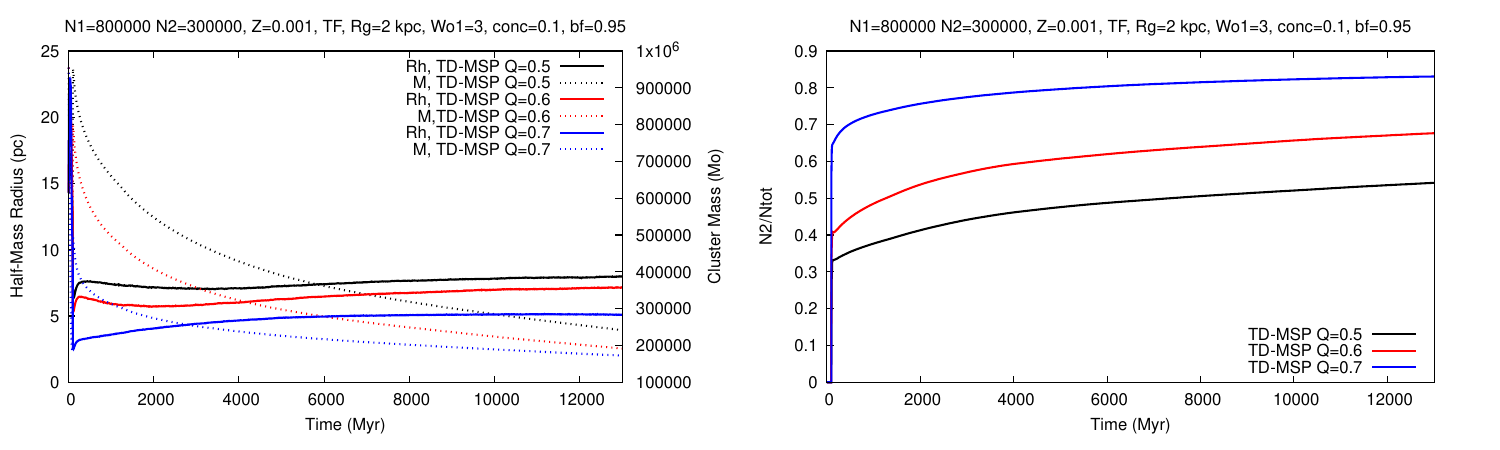} 
\caption{Left panel: Evolution of $\mathrm{R_h}$ and the cluster total mass for TD-MSP models with different $\mathrm{Q_1}$. Right panel: Evolution of the ratio $\mathrm{N_2/N_{tot}}$ for TD-MSP models with different $\mathrm{Q_1}$. The global cluster parameters are listed at the top of each panel and are the same as for Figure~\ref{f:Fig1}.}
\label{f:Fig3} 
\end{figure}

To achieve a stronger increase in the $\mathrm{N_2/N_{tot}}$ ratio after GC migration and to reproduce the observed values, an additional physical process is required. It is known that the removal of the residual gas after the formation of the \fg is associated with a strong expansion of the cluster and its departure from virial equilibrium \citep[e.g.,][]{Banerjee2013, BanerjeeKroupa2018}. We can therefore assume that the initial model for \fg has $\mathrm{Q_1 > 0.5}$. Increasing $\mathrm{Q_1}$ will lead to a decrease in the mass of the cluster and $\mathrm{R_h}$ and an increase in $\mathrm{N_2/N_{tot}}$, as is clearly seen in the Figure~\ref{f:Fig3}. The value of $\mathrm{Q_1}$ will depend on \fg star formation efficiency (SFE). The smaller SFE, the larger $\mathrm{Q_1}$. According to recent works, the amount of cluster expansion due to the residual gas removal is very uncertain and depends on many factors connected with the SFE, cluster structure, environment, and efficiency
of the feedback processes \citep[e.g.,][]{Farias2018, Geenetal2018, Zamora-Avilesetal2019, Krauseetal2020, Polaketal2024}. Therefore, the assumption that initially \fg is out of virial equilibrium and has $\mathrm{Q_1 > 0.5}$ is plausible.

\subsection{King parameter, $\mathrm{W_{o1}}$}\label{s:Wo}
\begin{figure}[h!]
\centering
\includegraphics[width=1.0\textwidth, height=0.24\textheight]{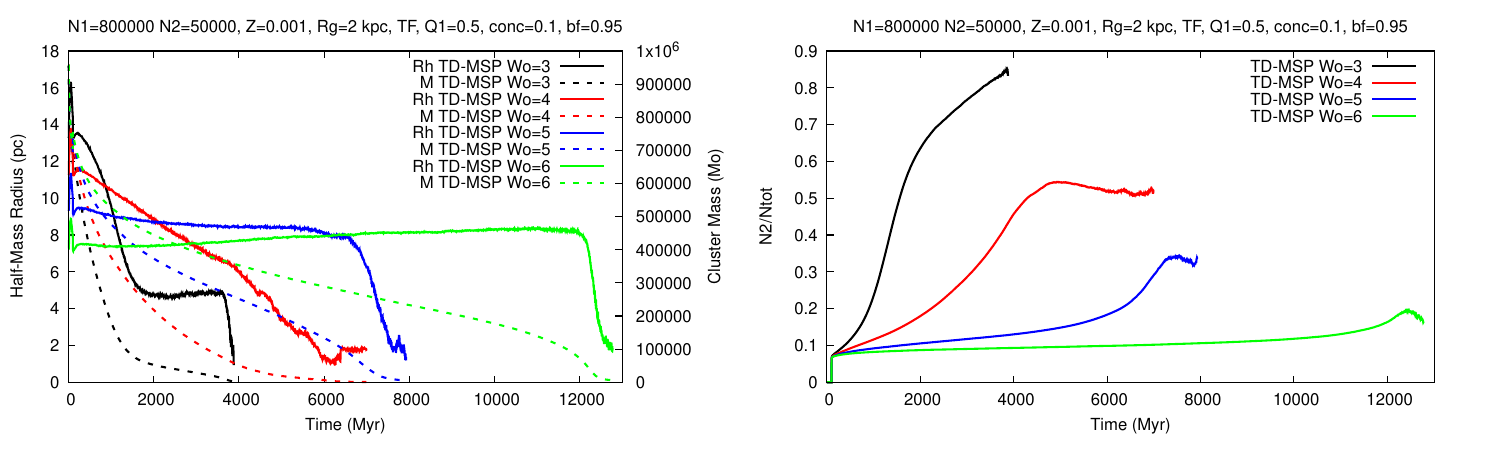}
    \caption{Left panel: Evolution of $\mathrm{R_h}$ and the cluster total mass for TD-MSP models with a different King parameter, $\mathrm{W_{o1}}$. Right panel: Evolution of the ratio $\mathrm{N_2/N_{tot}}$ for TD-MSP models with a different King parameter, $\mathrm{W_{o1}}$.  The global cluster parameters are listed at the top of each panel and are the same as for Figure~\ref{f:Fig1}.}
\label{f:Fig4} 
\end{figure}

The King parameter for \fg is crucial for the long-term evolution of the cluster and the $\mathrm{N_2/N_{tot}}$ ratio. As can be seen from Figure~\ref{f:Fig4} for $\mathrm{W_{o1}}$ larger than about 5, the $\mathrm{N_2/N_{tot}}$ ratio does not reach the observed values, and the models become similar to the single population models. The larger $\mathrm{W_{o1}},$ the larger GC evolution time scale, and the slower increase of $\mathrm{N_2/N_{tot}}$. This was already pointed out by \cite{Vesperinietal2021} and \cite{Hypki2022, Hypki2025}. A very interesting relationship can be observed in the right panel in Figure~\ref{f:Fig4}. For models with a small number of \sg stars relative to \fg stars, in the initial phase of cluster evolution, the $\mathrm{N_2/N_{tot}}$ ratio practically does not change, but after about 1-2 Gyr it increases rapidly and reaches the values observed for MW GCs. This feature of evolution may help to explain the correlation between the GC mass and the $\mathrm{N_2/N_{tot}}$ ratio.

\subsection{Galactocentric distance}\label{s:distance}
\begin{figure}[h!]
\centering
\includegraphics[width=1.0\textwidth, height=0.23\textheight]{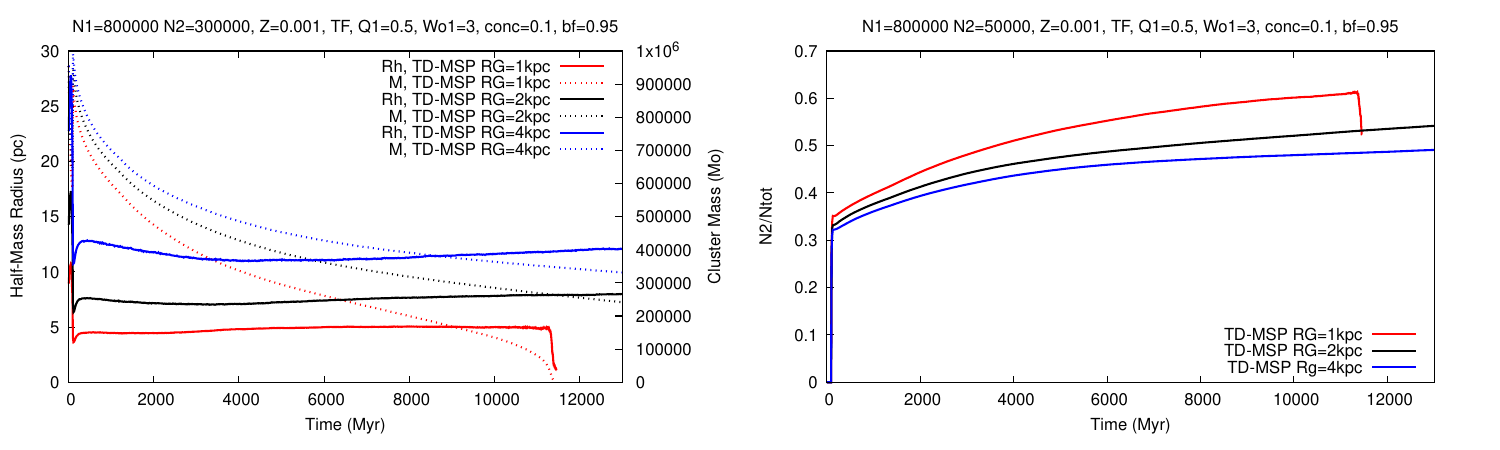} 
\caption{Left panel: Evolution of $\mathrm{R_h}$ and the cluster total mass for TD-MSP models with different galactocentric distance ($\mathrm{R_g}$). Right panel: Evolution of the ratio $\mathrm{N_2/N_{tot}}$ for TD-MSP models different ${R_g}$. The global cluster parameters are listed at the top of each panel and are the same as for Figure~\ref{f:Fig1}.}
\label{f:Fig5} 
\end{figure}

Another important parameter describing the evolution of MSP is $\mathrm{R_g}$ of the GC. As we can expect for the TF cluster, a larger distance of the GC from the galaxy center results in a larger tidal radius $\mathrm{R_t}$ and $\mathrm{R_h}$ and, consequently, a slower evolution of the cluster. This leads to the dependence, the larger $\mathrm{R_g}$ the larger $\mathrm{R_h}$ and the smaller $\mathrm{N_2/N_{tot}}$. Indeed, we can observe this type of evolution of global cluster parameters in Figure~\ref{f:Fig5}. The cluster with the smallest $\mathrm{R_g}$ is characterized by the smallest $\mathrm{R_h}$, mass and evolution time, and the largest $\mathrm{N_2/N_{tot}}$ ratio. It is important to draw attention to the fact that dependence on $\mathrm{R_g}$ is important from the point of view of MWGC observational parameters. The correlation between $\mathrm{R_g}$ and $\mathrm{N_2/N_{tot}}$ is not observed for MWGCs, and it arises from the specific initial conditions adopted in our models. This trend
is expected to be strongly blurred – or even erased – during the subsequent evolution of clusters due to environmental effects associated with galaxy assembly.

\subsection{Summary of the MOCCA simulations}\label{s:sim}
After a brief presentation of the results obtained with the MOCCA code for models with MSP with delayed \sg formation, it is time to summarize the results and put them in a broader perspective. 

The masses of MWGCs are contained in the range of roughly $\mathrm{10^4}$ to $\mathrm{10^6} M_{\odot}$, the $\mathrm{R_h}$ in the range of about 1 to 10 pc (larger radii on average have ex-situ GCs), and the $\mathrm{N_2/N_{tot}}$ ratio in the range of about 0.3 to 0.9. An exceptionally strong constraint on models of MSP formation in GCs is the correlation between the mass of the cluster and the $\mathrm{N_2/N_{tot}}$ ratio. The MOCCA simulations provide the following suggestions that may help understand the observational parameters of GCs with MSP in the framework of the AGB scenario:
\begin{itemize}
\item GCs just after the ejection of the residual gas should be TF or only slightly nTF. This means that they should preferentially form close to the galactic center. Too large $\mathrm{R_g}$ means much larger $\mathrm{R_t}$, and it is difficult in this case to form a cluster that is TF;
\item Only models in which the GC is TF or only slightly nTF reproduce $\mathrm{N_2/N_{tot}}$ ratios within the range observed for Milky Way GCs, provided that initially this ratio was not as large as observed;
\item For newly formed GCs of the same mass, increasing $\mathrm{R_g}$ leads to larger $\mathrm{R_h}$ and a lower $\mathrm{N_2/N_{tot}}$;
\item Models with smaller $\mathrm{W_{o1}}$ guarantee larger values of the $\mathrm{N_2/N_{tot}}$ ratio. The larger the $\mathrm{W_{o1}}$ the greater the mass and $\mathrm{R_h}$, but too small $\mathrm{W_{o1}}$ will lead to too fast cluster dissolution;
\item The greater the cluster's departure from virial equilibrium, the greater the $\mathrm{N_2/N_{tot}}$ ratio. The larger the $\mathrm{Q_1}$ the smaller the mass and $\mathrm{R_h}$;
\item Migration of the cluster to larger $\mathrm{R_g}$ leads to an increase in present-day cluster mass and $\mathrm{R_h}$. The $\mathrm{N_2/N_{tot}}$ ratio remains virtually constant after migration.
\end{itemize} 

The reaction of the observational parameters to the changes in the global and environmental parameters discussed in the above points is summarized in Table~\ref{t:Tab1}.

\begin{table}[h!]
\centering
\caption{Response of the observational global cluster parameters to changes in the models of the most important parameters describing the cluster environment and MSPs.
The global cluster parameters are: cluster mass, $R_h$, and $N_2/N_{tot}$. The parameters describing the cluster model and MSPs are: $\mathrm{R_g}$, $\mathrm{W_{o1}}$, $\mathrm{Q_1}$ and migration. ↑ means increase, ↓ means decrease, and = means unchanged.}
\begin{tabular}{cccc}
\hline\hline
Parameter              & M & $\mathrm{R_h}$ & $\mathrm{N_2/N_{tot}}$ \\ \hline
$\mathrm{R_g}$~$\uparrow$                  & = & $\uparrow$  & $\downarrow$       \\
$\mathrm{W_{o1}}$~$\uparrow$       & $\uparrow$ & $\uparrow$  & $\downarrow$    \\
$\mathrm{Q_1}$~$\uparrow$        & $\downarrow$ & $\downarrow$  & $\uparrow $  \\
migration~$\uparrow$      & $\uparrow$ & $\uparrow$  & =   \\  \hline 
\end{tabular}
\label{t:Tab1}
\end{table}

When talking about modeling MWGC, the main focus should be on in-situ GCs, because ex-situ GCs were formed in other environments. GCs are formed in the very early stages of the galaxy's evolution, when the galaxy is a very hostile environment, with a highly changing tidal field and multiple close and strong interactions with nearby dwarf galaxies. The mass of the galaxy is rapidly being built up during this period. The work of \cite{Meng2022} suggests that the period of rapid changes in the tidal field of a galaxy ends after about 1 Gyr. After this time, GCs virtually stop migrating outside. As many numerical simulations \citep[e.g.,][]{Banerjee2013, Levequeetal2022} have shown, gas ejection after the formation of \fg is associated with strong cluster expansion and excess of kinetic energy relative to potential energy ($\mathrm{Q_1 > 0.5}$). Such a cluster will be characterized by a relatively low density contrast (rather small $\mathrm{W_{o1}}$, significantly less than 5 or 6), and for small galactocentric distances, it should be TF. The cluster expansion due to gas expulsion is mainly related to SFE and not to the position of the cluster in the galaxy. Thus, clusters that form closer to the center have smaller $\mathrm{R_t}$ than those that arise further away, and can also more easily become TF or even TF overfilling, even if the SFE may be high as suggested by recent works by \cite{Caluraetal2019, Polaketal2024}.

The joint action of cluster migration and initial models with $\mathrm{Q_1 > 0.5}$ seems to work in the desired direction - $\mathrm{R_h}$ of a few pc and significant values of $\mathrm{N_2/N_{tot}}$, provided that the cluster is formed relatively close to the galactic center and it is initially relatively massive, at least about $\mathrm{10^6 M_{\odot}}$. 

The initial model parameters for the MOCCA simulations were chosen to explore the impact of specific assumptions related to MSP formation and evolution – rather than to reproduce the full diversity of the MWGC population – the agreement between the simulation results and observations is generally satisfactory, particularly in terms of the $\mathrm{N_2/N_{tot}}$ values. However, we acknowledge that the simulated clusters exhibit an anticorrelation between $\mathrm{N_2/N_{tot}}$ and present-day cluster mass, whereas observational data suggest a positive correlation. This discrepancy will be further discussed in the next Section, where a speculative scenario for MSP formation will be presented.

\section{Speculative AGB scenario for MSP formation}\label{s:specula}
Based on the results of the MOCCA simulations and the available literature on GC formation processes in zoom-in cosmological and purely hydrodynamical simulations, we introduce a speculative extension of the AGB scenario to provide a sketch of possible GC evolution in the strongly and rapidly changing environment of a Galaxy in the process of formation.  To reproduce the mass-$\ mathrm{N_2/N_{tot}}$ correlation, let’s assume that the availability of gas from which GCs can form decreases with distance from the galactic center. Thus, clusters forming farther from the center will have, on average, a lower mass of \fg and, more importantly, probably a significantly lower mass of gas that can again be re-accreted into the cluster and form \sg. The small value of $\mathrm{N_2}$ relative to $\mathrm{N_1}$ causes the increase in the $\mathrm{N_2/N_{tot}}$ ratio to be slow initially and only increase significantly after a period of 1 - 2 Gyr. Thus, if the cluster migrates during this time, its observed value of the $\mathrm{N_2/N_{tot}}$ ratio will be relatively low. For massive clusters formed from very massive gas clouds, the availability of gas is very high, which will lead to the formation of clusters that initially have a relatively large $\mathrm{N_2/N_{tot}}$ ratio. For such clusters, the increase in this ratio to large values occurs quickly and then stabilizes. This means that their migration can stop the $\mathrm{N_2/N_{tot}}$ratio at relatively large values. Dynamical friction also seems to be an important player in this process. We know that its effectiveness depends on the mass of the cluster and the distance from the center of the galaxy. Therefore, very massive clusters formed too close to the center or located after migration in not very eccentric orbits will sink and merge in the galactic nucleus. Those that have larger eccentricities have a larger probability of surviving. Less massive clusters are likely to be born or to migrate to larger distances, and therefore, dynamical friction will not be effective for them. Therefore, the mutual interaction of dynamical friction, migration, and availability of gas from which GC can be formed may lead to the generation of the currently observed correlation between the mass of GCs and their $\mathrm{N_2/N_{tot}}$ ratio. 

This is a rather speculative and perhaps a bit naive scenario requiring several physical processes to operate in the right time order and in the right environment of a young, assembling galaxy. Maybe this is the answer to the question of why, for young and massive star clusters that we are now observing, we do not see the observational signatures of MSPs. Simply, the current galactic environment is not suitable for the physical processes that enable the creation of MSP. 

The scenario presented above should not be taken as the only explanation for the observed properties of MSPs in MWGCs. In our view, it is the leading scenario and accounts for the most important features of MSPs. However, other scenarios discussed in the literature may also contribute, operating alongside the AGB scenario and further blurring the photometric and spectroscopic distinctions between \fg and \sg stars. Indeed, it would be surprising if these additional mechanisms did not play some role in shaping what we observe in MWGCs.

Finally, we note that according to \cite{Ishchenkoetal2024}, the orbits of MWGCs can undergo significant evolution over a Hubble time due to variations in the Galactic tidal field, resulting in either inward or outward migration. 
%This migration further blurs the initial parameters of the clusters and introduces a large scatter in their kinematic observational parameter.

\section{Discussion}\label{s:disc}
A good theory is always able to explain new observational facts. Let us then check whether the proposed speculative extension of the AGB scenario is able to explain the observations that have caused much trouble for all the MSP formation scenarios found in the literature.

\begin{figure}[h!]
\begin{center}
\includegraphics[width=0.85\linewidth]{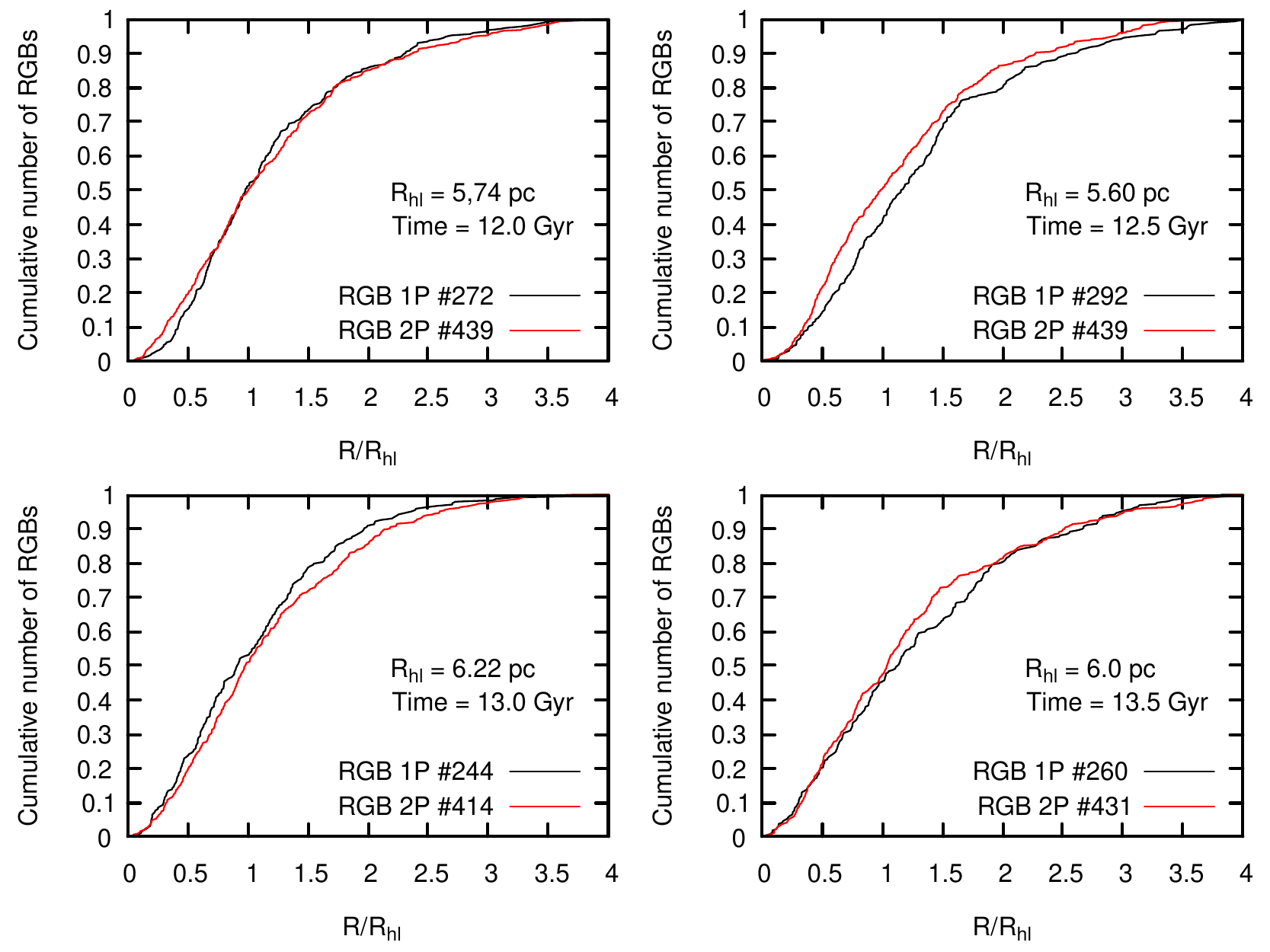}
\caption{Cumulative number distributions of RGB stars for \fg and \sg as a function of distance scaled by half-light radius ($R_{hl}$) for selected projected (2D) snapshots from \rm{12} Gyr to \rm{13.5} Gyr. \fg is more centrally concentrated than \sg only in the snapshot at \rm{13} Gyr. The number distributions of both populations change significantly over time, and the overconcentration of \fg is a transient feature. The $R_{hl}$ values, snapshot times, and the number of RGB stars are provided in the insets of each panel.
}
\label{f:Fig6}
\end{center}
\end{figure}

\cite{Leitingeretal2023} and later \cite{Cadelano2024} published papers in which they showed that for some clusters (namely, NGC 3201 and NGC 6101) the A+ parameter indicates that \fg is more concentrated than 2P. Virtually none of the proposed scenarios can explain these observations.  
Inspired by Leitinger's results, we analyzed the MOCCA simulation results and found many examples where \fg is more concentrated than 2P. All these models are characterized by a not very large cluster mass at present and contain a BH subsystem. The clusters observed by \cite{Leitingeretal2023} are clusters captured by the MW during merger events with dwarf galaxies. This means that these clusters, despite their small mass, have large tidal radii, and they are dynamically old - they are located at a large distance from the galactic center. To make the MOCCA models similar to the observed NGC 3201 and NGC 6101 clusters, we assume that the GC forms in a dwarf galaxy close to its center, to be later captured by the MW to a distant orbit \citep[for details see]{Giersz2025b}. In this way, the tidal radius and the radius containing half the luminosity are close to the observed values. One such MOCCA model is presented in Figure~\ref{f:Fig6}.

In the Figure~\ref{f:Fig6} we see that overconcentration, calculated as in observations for red giant branch (RGB) stars, is a transient feature. It appears and disappears. It is important to stress that the overconcentration is only visible for RGB stars, for other types of stars, in particular main-sequence (MS) stars, it is not visible. Both populations of MS stars are completely mixed with each other. To better understand why only for RGB stars overconcentration is observed and transient, we analyzed the interactions of RGB stars during the evolution of the cluster - the details and results of the analysis can be found in \cite{Giersz2025b}. In short, the BH subsystem formed by \fg stars governs the cluster's evolution. Since \sg stars are more centrally concentrated, their RGB progenitor stars interact more strongly than those of 1P. As a result, they are more efficiently displaced from the cluster center or even escape, ultimately leading to an overconcentration of \fg RGB stars at the present day. A key requirement is that the number of RGB stars in the system is relatively small. Then, small-N statistics together with dynamical interactions can lead to \fg overconcentration. For larger N, the effect of stronger \sg RGB interaction is blurred. 
When observations are limited to RGB stars, small-number statistics and dynamical interactions can distort their spatial distribution, leading to biased conclusions about the overall distribution of MSPs. 

\section{Conclusions}\label{s:concl}
Based on about 200 MOCCA simulations, we analyzed the influence of the most important parameters describing the MSP and GCs on the evolution of the cluster and its observational properties after 12 Gyr of evolution. The aim of the study was to determine such parameter ranges so that the cluster model after 12 Gyr of evolution best matches the ranges of observational parameters of the MWGCs; in particular, the cluster mass, $\mathrm{R_h}$, and the $\mathrm{N_2/N_{tot}}$ ratio. Based on the simulation analysis and the results of cosmological zoom-in simulations, a speculative refinement of the AGB scenario for the formation and evolution of GCs with MSP was presented to obtain an observational correlation between cluster mass and the $\mathrm{N_2/N_{tot}}$ ratio. To achieve this, one must combine the influence of the strongly changing MW environment in which the cluster evolves with its internal evolution and the effects related to the
ejection of the residual gas after \fg formation. All those physical processes have to operate in the right time order and in the right environment of the galaxy that is just forming and assembling. In order to confirm this speculative
refinement, more sophisticated modeling of the evolving
tidal field in which the GCs form and evolve is required. 

It is very important to emphasize that models of GCs with
MSPs based on the AGB scenario require completely different
initial properties than models of clusters with a single population. Instead of GCs being highly concentrated and lying deep inside the Roche lobe, models that fill the Roche lobe are required. This carries strong constraints on where in the galaxy GCs are formed.

Our study suggests that the spatial distribution and, potentially, the kinematic properties of MSP may depend on the type of stars observed. This effect is particularly relevant for GCs with present-day masses of a few $\mathrm{10^5M_{\odot}}$, which have retained only 10–20\% of their initial mass. Despite the significant mass loss, such clusters may appear dynamically young due to BHS heating and migration to larger Galactocentric distances. When observations are limited to RGB stars, small-number statistics and dynamical interactions can distort their spatial distribution, leading to biased conclusions about the overall distribution of MSPs. The overconcentration of \fg RGB stars relative to \sg RGB stars appears to be a transient effect. Confirming the findings of \cite{Leitingeretal2023} and \cite{Cadelano2024}, that in some MW GCs, \fg RGB stars are more centrally concentrated than \sg RGB stars, requires observations of MS stars to see if the same trend is seen for them. Our MOCCA simulations show no \fg overconcentration relative to \sg among MS stars. Verifying this observationally would be a crucial step toward understanding MSP properties and supporting the AGB scenario.

\begin{acknowledgements}
MG, AH, GW, and LH were supported by the Polish National Science Center (NCN) through the grant 2021/41/B/ST9/01191. AA acknowledges support for this paper from project No. 2021/43/P/ST9/03167 co-funded by the Polish National Science Center (NCN) and the European Union Framework Programme for Research and Innovation Horizon 2020 under the Marie Skłodowska-Curie grant agreement No. 945339. AH acknowledges support by the IDUB grant 140/04/POB4/0006 at Adam Mickiewicz University in Poznan, Poland. The figures presented in the article come from \cite{Giersz2025a, Giersz2025b} published in the journal Astronomy \& Astrophysics under the CC-BY license.
\end{acknowledgements}

\bibliographystyle{iaulike}
\bibliography{ref.bib}

\begin{thebibliography}{}

\bibitem[{Banerjee} and {Kroupa}, 2013]{Banerjee2013}
{Banerjee}, S. \& {Kroupa}, P. 2013, {Did the Infant R136 and NGC 3603 Clusters
  Undergo Residual Gas Expulsion?}
\newblock {\em Astrophysical Journal}, 764(1), 29.

\bibitem[{Banerjee} and {Kroupa}, 2018]{BanerjeeKroupa2018}
{Banerjee}, S. \& {Kroupa}, P.
\newblock {Formation of Very Young Massive Clusters and Implications for
  Globular Clusters}.
\newblock In {Stahler}, S., editor, {\em The Birth of Star Clusters} 2018,,
  volume 424 of {\em Astrophysics and Space Science Library},  143.

\bibitem[{Bastian} and {Lardo}, 2018]{Bastian2018}
{Bastian}, N. \& {Lardo}, C. 2018, {Multiple Stellar Populations in Globular
  Clusters}.
\newblock {\em Annual Review of Astronomy \& Astrophysics}, 56, 83--136.

\bibitem[{Cadelano} et~al., 2024]{Cadelano2024}
{Cadelano}, M., {Dalessandro}, E., \& {Vesperini}, E. 2024, {The structural
  properties of multiple populations in globular clusters: The instructive case
  of NGC 3201}.
\newblock {\em Astronomy \& Astrophysics}, 685, A158.

\bibitem[{Calura} et~al., 2019]{Caluraetal2019}
{Calura}, F., {D'Ercole}, A., {Vesperini}, E., {Vanzella}, E., \& {Sollima}, A.
  2019, {Formation of second-generation stars in globular clusters}.
\newblock {\em Monthly Notices of the RAS}, 489(3), 3269--3284.

\bibitem[{Farias} et~al., 2018]{Farias2018}
{Farias}, J.~P., {Fellhauer}, M., {Smith}, R., {Dom{\'\i}nguez}, R., \&
  {Dabringhausen}, J. 2018, {Gas expulsion in highly substructured embedded
  star clusters}.
\newblock {\em Monthly Notices of the RAS}, 476(4), 5341--5357.

\bibitem[{Geen} et~al., 2018]{Geenetal2018}
{Geen}, S., {Watson}, S.~K., {Rosdahl}, J., {Bieri}, R., {Klessen}, R.~S., \&
  {Hennebelle}, P. 2018, {On the indeterministic nature of star formation on
  the cloud scale}.
\newblock {\em Monthly Notices of the RAS}, 481(2), 2548--2569.

\bibitem[{Giersz} et~al., 2025a]{Giersz2025a}
{Giersz}, M., {Askar}, A., {Hypki}, A., {Hong}, J., {Wiktorowicz}, G., \&
  {Hellstr{\"o}m}, L. 2025,a {MOCCA: Effects of pristine gas accretion and
  cluster migration on globular cluster evolution, global parameters, and
  multiple stellar populations}.
\newblock {\em Astronomy \& Astrophysics}, 699a, A76.

\bibitem[{Giersz} et~al., 2025b]{Giersz2025b}
{Giersz}, M., {Askar}, A., {Hypki}, A., {Hong}, J., {Wiktorowicz}, G., \&
  {Hellstr{\"o}m}, L. 2025,b {Multiple stellar populations in MOCCA globular
  cluster models: Transient spatial overconcentration of pristine red giant
  stars driven by strong dynamical encounters}.
\newblock {\em Astronomy \& Astrophysics}, 698b, L11.

\bibitem[{Gratton} et~al., 2019]{Gratton2019}
{Gratton}, R., {Bragaglia}, A., {Carretta}, E., {D'Orazi}, V., {Lucatello}, S.,
  \& {Sollima}, A. 2019, {What is a globular cluster? An observational
  perspective}.
\newblock {\em The Astronomy and Astrophysics Review}, 27(1), 8.

\bibitem[{Hypki} et~al., 2022]{Hypki2022}
{Hypki}, A., {Giersz}, M., {Hong}, J., {Leveque}, A., {Askar}, A., {Belloni},
  D., \& {Otulakowska-Hypka}, M. 2022, {MOCCA: dynamics and evolution of single
  and binary stars of multiple stellar populations in tidally filling and
  underfilling globular star clusters}.
\newblock {\em Monthly Notices of the RAS}, 517(4), 4768--4787.

\bibitem[{Hypki} et~al., 2025]{Hypki2025}
{Hypki}, A., {Vesperini}, E., {Giersz}, M., {Hong}, J., {Askar}, A.,
  {Otulakowska-Hypka}, M., {Hellstrom}, L., \& {Wiktorowicz}, G. 2025, {MOCCA:
  Global properties of tidally filling and underfilling globular star clusters
  with multiple stellar populations}.
\newblock {\em Astronomy \& Astrophysics}, 693, A41.

\bibitem[{Ishchenko} et~al., 2024]{Ishchenkoetal2024}
{Ishchenko}, M., {Berczik}, P., {Panamarev}, T., {Kuvatova}, D., {Kalambay},
  M., {Gluchshenko}, A., {Veles}, O., {Sobolenko}, M., {Sobodar}, O., \&
  {Omarov}, C. 2024, {Dynamical evolution of Milky Way globular clusters on the
  cosmological timescale: I. Mass loss and interaction with the nuclear star
  cluster}.
\newblock {\em Astronomy \& Astrophysics}, 689, A178.

\bibitem[{Krause} et~al., 2020]{Krauseetal2020}
{Krause}, M. G.~H., {Offner}, S. S.~R., {Charbonnel}, C., {Gieles}, M.,
  {Klessen}, R.~S., {V{\'a}zquez-Semadeni}, E., {Ballesteros-Paredes}, J.,
  {Girichidis}, P., {Kruijssen}, J.~M.~D., {Ward}, J.~L., \& {Zinnecker}, H.
  2020, {The Physics of Star Cluster Formation and Evolution}.
\newblock {\em Space Science Reviews}, 216(4), 64.

\bibitem[{Leitinger} et~al., 2023]{Leitingeretal2023}
{Leitinger}, E., {Baumgardt}, H., {Cabrera-Ziri}, I., {Hilker}, M., \&
  {Pancino}, E. 2023, {A wide-field view on multiple stellar populations in 28
  Milky Way globular clusters}.
\newblock {\em Monthly Notices of the RAS}, 520(1), 1456--1480.

\bibitem[{Leveque} et~al., 2022]{Levequeetal2022}
{Leveque}, A., {Giersz}, M., {Banerjee}, S., {Vesperini}, E., {Hong}, J., \&
  {Portegies Zwart}, S. 2022, {A Monte Carlo study of early gas expulsion and
  evolution of star clusters: new simulations with the MOCCA code in the AMUSE
  framework}.
\newblock {\em Monthly Notices of the RAS}, 514(4), 5739--5750.

\bibitem[{Meng} and {Gnedin}, 2022]{Meng2022}
{Meng}, X. \& {Gnedin}, O.~Y. 2022, {Tidal disruption of star clusters in
  galaxy formation simulations}.
\newblock {\em Monthly Notices of the RAS}, 515(1), 1065--1077.

\bibitem[{Polak} et~al., 2024]{Polaketal2024}
{Polak}, B., {Mac Low}, M.-M., {Klessen}, R.~S., {Wei Teh}, J.,
  {Cournoyer-Cloutier}, C., {Andersson}, E.~P., {Appel}, S.~M., {Tran}, A.,
  {Lewis}, S.~C., {Wilhelm}, M. J.~C., {Portegies Zwart}, S., {Glover}, S.
  C.~O., {Rieder}, S., {Wang}, L., \& {McMillan}, S. L.~W. 2024, {Massive star
  cluster formation: I. High star formation efficiency while resolving feedback
  of individual stars}.
\newblock {\em Astronomy \& Astrophysics}, 690, A94.

\bibitem[{Vesperini} et~al., 2021]{Vesperinietal2021}
{Vesperini}, E., {Hong}, J., {Giersz}, M., \& {Hypki}, A. 2021, {Dynamical
  evolution of multiple-population globular clusters}.
\newblock {\em Monthly Notices of the RAS}, 502(3), 4290--4304.

\bibitem[{Zamora-Avil{\'e}s} et~al., 2019]{Zamora-Avilesetal2019}
{Zamora-Avil{\'e}s}, M., {Ballesteros-Paredes}, J., {Hern{\'a}ndez}, J.,
  {Rom{\'a}n-Z{\'u}{\~n}iga}, C., {Lora}, V., \& {Kounkel}, M. 2019,
  {Flipping-up the field: gravitational feedback as a mechanism for young
  clusters dispersal}.
\newblock {\em Monthly Notices of the RAS}, 488(3), 3406--3415.

\end{thebibliography}

\end{document}